\documentclass{article}
\usepackage{misc/arxiv3}

\usepackage{amsmath}
\usepackage{url} 
\usepackage{amsthm} 
\usepackage[colorlinks=true,linkcolor=red,filecolor=green,citecolor=blue]{hyperref}
\usepackage{comment}
\usepackage{graphicx} 
\usepackage{caption}
\usepackage{subcaption}
\usepackage{algorithmic} 
\usepackage{url}
\PassOptionsToPackage{hyphens}{url}
\usepackage{hyperref}
\usepackage{bbm}
\expandafter\def\expandafter\UrlBreaks\expandafter{\UrlBreaks\do\-\do\~\do\'\do\"\do\-}%

\usepackage{changepage}
\usepackage{appendix}

\usepackage[dvipsnames]{xcolor} 

\definecolor{colour1}{RGB}{166,206,227}
\definecolor{colour2}{RGB}{31,120,180}
\definecolor{colour3}{RGB}{178,55,250} 
\definecolor{colour4}{RGB}{51,160,44}

\RequirePackage[authoryear]{natbib}
\title{A Bayesian mixture model for Poisson
network autoregression}

\author[1]{Elly Hung}
\author[1]{Anastasia Mantziou}
\author[2]{Gesine Reinert}
\affil[1]{Department of Statistics, University of Warwick, Coventry, CV4 7AL, UK}
\affil[2]{Department of Statistics, University of Oxford, 24-29 St Giles, Oxford OX1 3LB, UK}

\date{}

\begin{document}

\maketitle

\section{Introduction}
Multivariate count time series arise in a wide range of applications, including the number of COVID-19 cases recorded each week in different counties. While for this example, in \cite{Armbruster24} a model is used that assumes Gaussian errors, for count data this assumption is not natural. Moreover, the model does not reflect any clustering in the nodes of the network which may affect the process. With this motivating example, we develop a model with the following aims. We assume that the time series occur on the nodes of a known underlying network where the edges dictate the form of a structural vector autoregression model. In contrast to using a full vector autoregressive model, the network assumption is a means of imposing sparsity, as  for example in \cite{Armillotta2023}. A second aim is to accommodate heterogeneous node dynamics, and to develop a probabilistic model for clustering nodes that exhibit similar behaviour. To address these aims, we propose a new Bayesian Poisson network autoregression mixture model that we call PNARM, which combines ideas from the models of \cite{Armillotta2023}, \cite{Dahl2008}, and \cite{Ren2024}.

\section{Method}
The PNARM model assumes that nodes have a latent class label which affects its autoregressive properties. The class assignment itself is random as well. Mathematically,
let $Z_i$ denote the latent class of node $i$, $S(\textbf{Z})$ denote a partition induced by the cluster labels $\textbf{Z}$, with $S(\textbf{Z})^{(-i)}$ being the partition with node $i$ removed, and let $\mathbf{Z} = (Z_1, Z_2, ..., Z_N)$, $K(\mathbf{Z}) = \max_i Z_i$ so that $Z_i \in \{1, .., K(\mathbf{Z})\}$. Let $\mathbf{Y_i}$ be the length $T$ time series observed on node $i$, and let $\mathbf{Y_{-i}}$ be the time series observed on all the other nodes of the network. 
The PNARM model is given in hierarchical form as
\begin{equation}\label{eq:GPAR}
\begin{split}
    \mathbf{Z} &\sim \pi_\mathbf{Z} \\
    \mathbf{\theta_k} &\stackrel{\text{i.i.d.}}{\sim} \pi_\theta \text{ for } k = 1, \dots, K(\mathbf{Z})\\
p(\mathbf{Y_i} | \mathbf{Z, \theta, {Y_{-i}}}) &\propto \prod_{t=2}^T \text{Poisson}(Y_{i, t}; \lambda_{i, t, \mathbf{\theta_{Z_i}}} := \theta_{1, Z_i} v_i + \theta_{2, Z_i} X_{i, t-1} + \theta_{3, Z_i} Y_{i, t-1})
\end{split}
\end{equation}
where $\pi_\mathbf{Z}, \pi_\mathbf{\theta}$ denote the priors for the partition and the autoregression coefficients, and $\lambda_{i, t, \theta}$ is the expected value of the observation for node $i$ at time $t$ conditional on $\theta$ and on the process up to time $t-1$. 
The PNARM model extends the graphical assistant grouped network autoregression model (GAGNAR) \citep{Ren2024} from a Gaussian to a Poisson distribution and generalises the prior for the partition. 
The linear Poisson network autoregression (PNAR(1)) model by \cite{Armillotta2023} 
is a frequentist special case of \eqref{eq:GPAR}. 
Instead of using the PNAR(1) predictors, another approach that may help with modelling heterogeneous node dynamics is to have $v_i$ and $X_{i, t-1}$ depend on known network information. For example, to account for the difference in population sizes across different counties, we consider applying a population adjustment, taking the predictors to be $v_i = \frac{p_i}{c} \text{ and } X_{i, t-1} = \frac{p_i}{deg(i)} \sum_{j=1}^N A_{i, j} \frac{1}{p_j} Y_{j, t-1}$, where $p_i$ is the population in county $i$ and $c \in \mathbb{R}_{>0}$ is some fixed constant.

We investigate two possible choices of partition prior $\pi_\mathbf{Z}$: the Dirichlet-multinomial finite mixture model (FMM) with a non-informative prior for mixture proportions, and a distant-dependent partition prior (DDP) by \cite{Dahl2008} which generalises the uniform co-clustering assumption in the Dirichlet process, by implicitly defining a prior through the co-clustering probabilities 
\begin{equation} \label{eq:Dahl_prior}
p(i \in S_k | S^{(-i)}) \propto
\begin{cases}
    \sum_{j: j \neq i, \,j \in S_k^{(-i)}} h_{ij}, & k = 1, ..., K(S^{(-i)}) \\
    \alpha, & k=K(S^{(-i)})+1,
\end{cases} 
\end{equation}
where the weights $h_{ij}$ are scaled to satisfy $\sum_{j: j \neq i} h_{ij} = N-1$. Hence, the probability of forming a new cluster is $\frac{\alpha}{\alpha + N- 1}$, the same as in a Dirichlet process mixture model -- setting $h_{ij} = 1$ for all $i, j$ recovers the Dirichlet process prior. We set the co-clustering weights between $i$ and $j$ to be  $h_{ij} \propto \exp(-h \times d_{ij})$, where $d_{ij}$ is the shortest path between nodes $i$ and $j$. This choice of $h_{ij}$ means that large values of $h$ decreases the prior probability of a node being clustered with its neighbours, exponentially down-weighted for nodes that are further away. As $\alpha$ increases, the expected number of clusters increases. 

To target samples from the posterior of \eqref{eq:GPAR}, we use a Markov Chain Monte Carlo (MCMC) algorithm that alternates between sampling cluster labels for the nodes via a Gibbs sampler and sampling coefficients for the clusters using a random-walk Metropolis step. To mitigate against invalid inference due to non-converging chains for the finite mixture model, we explored a similar chain stacking approach to \cite{Yao2022}, but replace the leave-one-out predictive distribution due to the dependencies in an autoregressive time series, and instead use the predictive distribution for time steps in a validation set. 

Let index the MCMC iteration, with $(\mathbf{Z}^{(m)}, \theta^{(m)})$ being the $m$-th sample, $M$ is the total number of MCMC samples, and $S^{(m)}$ is the partition induced by the cluster labels $\mathbf{Z}^{(m)}$. Let $\hat{c}_{ij} := \frac{1}{M} \sum_{m=1}^M \mathbbm{1}_{z_i^{(m)} = z_j^{(m)}}$ denote the proportion of times that node $i$ and $j$ are allocated to the same cluster. 
As in \cite{Dahl2008} we select a partition among the samples that minimises the loss function 
  $  S^{LS} := \min_{S^{(1)}, S^{(2)}, ..., S^{(M)}} \sum_{i \neq j} (\mathbbm{1}_{z_i^{(m)} = z_j^{(m)}} - \hat{c}_{ij})^2 .$ 
  
  While the exact posterior predictive distribution is intractable we can approximate it by
\begin{equation} \label{eq:predictive_distribution}
    \hat{p}(\mathbf{y_{T+1}} | \mathbf{y_{1:T}}) = \frac{1}{M} \prod_{i=1}^N \sum_{m=1}^M \text{Poisson}\left (y_{i, T+1}; \theta_{1, z_i^{(m)}}^{(m)} w_i + \theta_{2, z_i^{(m)}}^{(m)} x_{i, T} + \theta_{3, z_i^{(m)}}^{(m)} y_{i, T} \right).
\end{equation}

\section{Results on COVID-19 cases in Ireland}

\begin{figure}[htbp]
    \centering
    \begin{minipage}{0.32\textwidth}
        \centering
        \includegraphics[width=\linewidth]{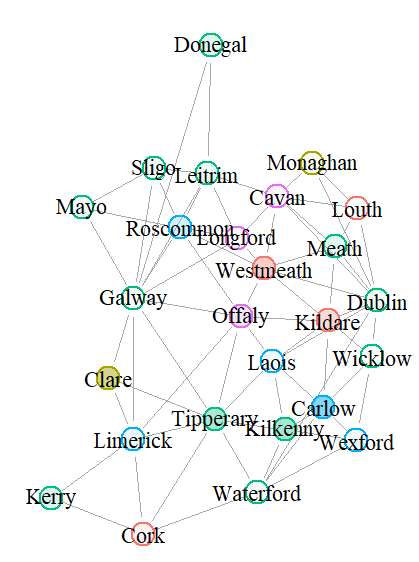}
        \subcaption{5 component FMM}
    \end{minipage}\hfill
    \begin{minipage}{0.32\textwidth}
        \centering
        \includegraphics[width=\linewidth]{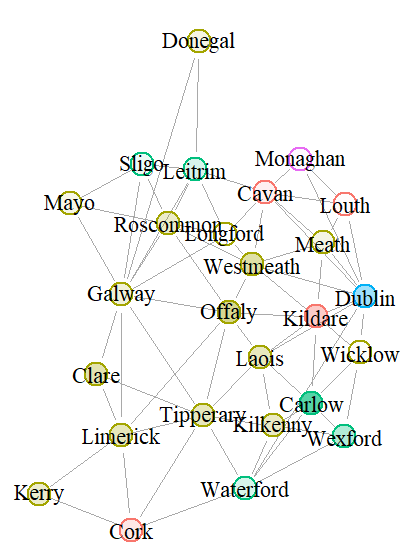}
        \subcaption{GAGNAR}
    \end{minipage}\hfill
    \begin{minipage}{0.32\textwidth}
        \centering \includegraphics[width=\linewidth]{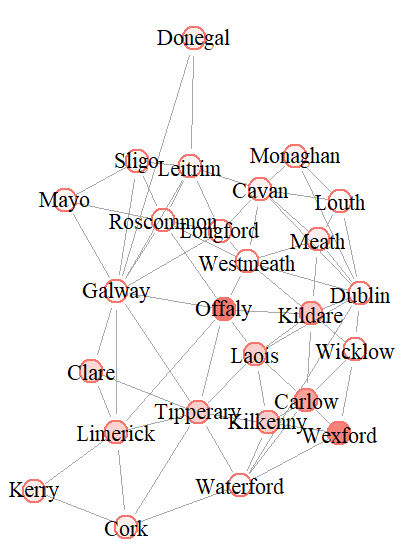}
        \subcaption{PNAR(1) with population-adjusted predictors}
    \end{minipage}
    \caption{Magnitudes of scaled errors from forecasts: colours correspond to least-squares partitions from the 5-component mixture model and GAGNAR models and higher opacity reflects higher scaled error in the point forecast. Note that there is no clustering for the  PNAR model.}
    \label{fig:cluster_error}
\end{figure}

As in \cite{Armbruster24}, we consider the weekly number of COVID-19 cases in the counties of the Republic of Ireland, from 01/03/2020 for 25 weeks. 
We use the economic hubs network from \cite{Armbruster24} as our underlying network, which is constructed by connecting counties that share borders, then add additional edges to connect each county to its nearest economic hub: Dublin, Cork, Limerick, Galway, or Waterford, as a proxy for commuter flow. Table \ref{tab:metrics} compares the predictive performance of the PNARM models to the GAGNAR and PNAR models, while Figure \ref{fig:cluster_error} gives a visualisation of the magnitudes of the mean absolute scaled errors, see  \cite{HYNDMAN2006}, across different counties and the clusters obtained under different models. There seems to be a correspondence between  counties with very large or very small ratios 
of the values observed in week 24 versus week 25 and counties with large prediction errors. From preliminary MCMC runs, other DDP hyperparameter values of $\alpha, h$ seemed to produce similar posterior distributions for the coefficients.

\begin{table}[h!] 
\centering
\begin{tabular}{|l|c|c|c|}
\hline
Model specification \,& Mean abs. scaled error  \,& Training score \,& Test score\,\\
\hline
PNARM, DDP prior, with $h=1$, $\alpha=1$ & 0.52 & \textbf{6.23} & \textbf{4.47}\\
PNARM, 5-component FMM & \textbf{0.46} & 6.42 & 4.81\\
PNARM, 4-component FMM & 0.50 & 6.53 & 5.09\\
GAGNAR & 0.65 & 6.99 & 6.99\\
PNAR with raw counts & 0.70 & 7.07 & 6.52\\
PNAR with population-adjusted predictors & 0.68 & 7.13 & 6.17\\
\hline 
\end{tabular}
\caption{Performance metrics of different models applied to the Ireland COVID-19 data. The PNARM models used a $\Gamma(1, 1)$ prior for the coefficients. The score refers to $\frac{1}{N |\mathcal{T}|}\sum_{i=1}^N \sum_{t \in \mathcal{T}} -\log (P^{(i, t)}(y_{i, t}) - P^{(i, t)}(y_{i, t} - 1))$, with the training data being $\mathcal{T}$=\{2, ..., 24\} and the test data $\mathcal{T}$=\{25\} and $P^{(i, t)}$ being the predictive cumulative distribution function \citep{czado2009predictive}. The model with the best metric is in \textbf{bold}.}
\label{tab:metrics}
\end{table}
\vspace{-1cm}

\section{Conclusion and future work}
Our initial work suggests that the PNARM(1) models outperform the GAGNAR(1) and PNAR(1) models when applied to the Ireland COVID-19 data set, showing the utility of having network time series models tailored for count distributions that allow for heterogeneous node behaviour. We observe that the Adjusted Rand Index comparing the least-squares partition obtained by the 5-component and GAGNAR models is 0.31; we conjecture that the differences in the partitions could arise because the GAGNAR model assumes constant innovation variance within a cluster, unlike the PNARM model. 

To assess calibration of the models, we computed randomised probability integral transforms (PIT), which under the true data-generating process, should be standard uniform. However, histograms of the randomised PITs of the training data exhibited U-shapes, suggesting overdispersion relative to the models with Poisson distribution, while the histogram for the GAGNAR model had PITs concentrated around the 0.5 region. Further work may look into other models for count distributions, such as a negative-binomial distribution, or a Poly\'a-Aeppli process to take into account the COVID-19 cases that occur as a cluster outbreak \citep{COVIDhpsc2020}. 

\medskip
{\bf Acknowledgement.} This work is supported in part by EPSRC grants EP/T018445/1 and EP/X002195/1.

\bibliography{sample} 

\end{document}